\documentclass[a4paper,11pt]{article}
\pdfoutput=1 

\usepackage{jcappub} 
	
\usepackage{lscape}
\usepackage[T1]{fontenc} 
\usepackage{bm}
\usepackage{mathrsfs}
\usepackage{aas_macros}
\usepackage{color}
\usepackage[dvipsnames]{xcolor}
\usepackage{multirow}
\usepackage{verbatim}
\usepackage{hyperref}
\newcommand{\eLISA}{{\em LISA}}
\newcommand{\TianQin}{{\em TianQin}}
\newcommand{\Taiji}{{\em Taiji}}
\newcommand{\Msun}{M_\odot}
\newcommand{\CEL}{CEL}
\newcommand{\mchirp}{M_{\rm chirp}}
\newcommand{\joseprd}[1]{\textcolor{black}{#1}} 

\title{\boldmath \emph{Chirping} compact stars: gravitational radiation and detection degeneracy with binaries
}

\author[a,c,1]{J.~F.~Rodr\'iguez,\note{Corresponding author.}}
\author[b,c,d,e,f]{J.~A.~Rueda,}
\author[b,c,g]{R.~Ruffini}
\author[h]{J.~I.~Zuluaga}
\author[i,j]{J.~M.~Blanco-Iglesias}
\author[i,j]{P.~Lor\'en-Aguilar}

\affiliation[a]{Escuela de F\'isica, Universidad Industrial de Santander, \\
Ciudad Universitaria, Bucaramanga 680002, Colombia}
\affiliation[b]{ICRA, Dip. di Fisica, Sapienza Universit\`a  di Roma, \\
Piazzale Aldo Moro 5, I-00185 Roma, Italy}
\affiliation[c]{ICRANet,
Piazza della Repubblica 10, I-65122 Pescara, Italy}
\affiliation[d]{ICRANet-Ferrara, Dip. di Fisica e Scienze della Terra, Universit\`a di Ferrara, \\
Via Saragat 1, I-44122 Ferrara, Italy}
\affiliation[e]{Dip. di Fisica e Scienze della Terra, Universit\`a di Ferrara,\\
Via Saragat 1, I-44122 Ferrara, Italy}
\affiliation[f]{INAF, Istituto di Astrofisica e Planetologia Spaziali, \\
Via Fosso del Cavaliere 100, 00133 Rome, Italy}
\affiliation[g]{INAF, Viale del Parco Mellini 84, I-00136 Rome, Italy}
\affiliation[h]{Solar, Earth and Planetary Physics Group (SEAP) Instituto de F\'isica - FCEN, Universidad de Antioquia,\\
Calle 67 No. 53--108, Medell\'in, Colombia}
\affiliation[i]{Departament de F\'isica, Universitat Polit\`ecnica de Catalunya, \\
c/Esteve Terrades, 5, 08860 Castelldefels, Spain
}
\affiliation[j]{School of Physics, University of Exeter, \\
Stocker Road, Exeter EX4 4QL, UK}

\emailAdd{joferoru@correo.uis.edu.co, jorge.rueda@icra.it, ruffini@icra.it,jorge.zuluaga@udea.edu.co, P.Loren-Aguilar@exeter.ac.uk, josemiguelblancoiglesias@gmail.com}

\abstract{Compressible, Riemann S-type ellipsoids can emit gravitational waves (GWs) with a chirp-like behavior (hereafter {\em chirping} ellipsoids, CELs). We show that the GW frequency-amplitude evolution of CELs (mass $\sim 1$~M$_\odot$, radius $\sim10^3$~km, polytropic equation of state with index $n\approx 3$) is indistinguishable from that emitted by double white dwarfs and by extreme mass-ratio inspirals (EMRIs) composed of an intermediate-mass (e.g.~$10^3~M_\odot$) black hole and a planet-like (e.g.~$10^{-4}~M_\odot$) companion, in the frequency interval within the detector sensitivity band in which the GW emission of these systems is quasi-monochromatic. 
For reasonable astrophysical assumptions, the local universe density rate of CELs, double white dwarfs, and EMRIs in the mass range here considered are very similar, posing a detection-degeneracy challenge for space-based GW detectors. We outline the astrophysical implications of this CEL-binary detection degeneracy by space-based GW-detection facilities.
}

\begin{document}
\maketitle
\flushbottom

\section{Introduction}\label{sec:1}

Space-based, gravitational wave (GW) interferometers such as \eLISA\ \cite{2017arXiv170200786A}, \TianQin\  \cite{2016CQGra..33c5010L} and \Taiji\  \cite{Hu:2017mde} have the potential to detect low-frequency GWs and thus to give details of a different set of astrophysical objects with respect to the ones detectable by Earth-borne interferometers such as LIGO/Virgo. Specifically, \eLISA\ is sensitive to the frequency range $10^{-5} - 1$~Hz \cite{2004PhRvD..69h2005B, 2012CQGra..29l4016A}, and \TianQin\ in the range $10^{-4} - 0.1$ Hz \cite{2016CQGra..33c5010L}.

Among the main astrophysical targets expected for these detectors are the so-called extreme mass-ratio inspirals (EMRIs), namely binaries with symmetric mass-ratios $\nu \equiv q/(1+q)^2\approx q \equiv m_2/m_1\ll 1$. EMRIs\- that fall within the aforementioned GW frequency range are, for example, binaries composed of a supermassive (e.g.~$m_1 \sim 10^6$--$10^9~M_\odot$) or intermediate-mass (e.g.~$m_1 \sim 10^2$--$10^3~M_\odot$) black hole, accompanied by a stellar-mass object (e.g.~$m_2\sim M_\odot$) or, more interestingly (for the purposes of the present work), a substellar object (e.g.~$m_2 \ll M_\odot$), respectively (see eg. \cite{2017PhRvD..95j3012B} and references therein). 

Another target for GW detectors is represented by triaxial objects (e.g., deformed compact stars) emitting gravitational radiation while approaching axial symmetry. Searches for GWs from deformed neutron stars have been conducted in LIGO/Virgo detectors in the Hz-kHz band (e.g.~\cite{2018PhRvD..97j2003A,2017ApJ...839...12A}). So far, no analogous sources in the sub-Hz frequency region appear to have been considered possible targets of \eLISA\, even for different types of stellar objects. These sources could help to test astrophysical and relativistic objects, such as white dwarfs and low-mass compact objects, in physical regimes not previously explored and with unprecedented precision.

We show in this work that:
\begin{enumerate}
\item 
Triaxial, white dwarf-like compact objects emit \joseprd{quasi-monochromatic} detectable GWs in the \eLISA\ frequency sensitivity band;

\item
Their GW emission (spectrum, spanned frequency range, and time evolution), \textcolor{black}{becomes almost indistinguishable from that of some binaries, specifically detached double white dwarfs and EMRIs in the case of intermediate-mass black holes with planet-like companions}.
\end{enumerate}

\textcolor{black}{
We aim to characterize the above detection degeneracy. This challenge makes difficult the unambiguous identification of these objects by space-based interferometers, also given their expected comparable rates.
}

\textcolor{black}{
The article is organized as follows. In Sec.~\ref{sec:2}, we recall the main physical properties of the compressible, triaxial ellipsoid-like object relevant for this work, which we name \emph{chirping ellipsoid} (\CEL). The properties of the GW emission from a \CEL\ are investigated in Sec.~\ref{sec:3}. We identify white dwarf-like objects as the kind of \CEL\ that could mimic the GW emission from some binaries. In Sec.~\ref{sec:4}, we summarize the main quantities relevant to estimating the detectability by space-based interferometers of the GWs from CELs. We also define when we can consider GWs monochromatic in the interferometer band. Having defined these key ingredients, we identify in Sec.~\ref{sec:5} the binary systems for which a detection degeneracy with \CEL\ occurs. We refer to them as \CEL\ equivalent binaries. Section~\ref{sec:6} is devoted to giving estimates of the rates of \CEL\ and the equivalent binaries. Finally, in Sec.~\ref{sec:7}, we draw our conclusions.
}

\section{Evolution of compressible ellipsoids}\label{sec:2}

The study of equilibrium configurations of rotating self-gravitating systems using analytic methods (e.g.,~\cite{chandrasekharellipsoidal}) allows us to estimate the gravitational radiation emitted by rotating stars. In \cite{1969ApJ...158L..71F,1970ApJ...161..561C,1970ApJ...161..571C,1974ApJ...187..609M} incompressible rotating stars were studied following a quasi-static evolution approach.
In particular, it was shown that Kelvin's circulation, $C\equiv\pi a_1 a_2\bigl(\zeta + 2 \Omega \bigr)$\footnote{Not to be confused with the compactness parameter denoted by \emph{calligraphic} $\mathcal{C}$.},
is conserved when the dynamics is only driven by gravitational radiation reaction \cite{1974ApJ...187..609M}. Hereafter, the principal axes of the ellipsoid are denoted by $(a_1, a_2, a_3)$, the angular velocity of rotation around $a_3$  by $\Omega$, and the vorticity in the same direction by $\zeta$. In \cite{1993ApJS...88..205L,1995ApJ...442..259L} it was studied the GW emission of compressible, rotating stars with matter described by a polytropic equation of state, i.e., $P = K \rho^{1 + 1/n}$, where $P$ is the pressure, $\rho$ the matter density, and $n$ and $K$ are the polytropic index and constant.

We are interested in the GW emission of Riemann type-S ellipsoids \cite{chandrasekharellipsoidal}, which are not axially symmetric but whose equilibrium sequences of constant circulation can be constructed.

There are two main sequences of rotating triaxial ellipsoids: the \emph{Jacobi-like} (spinning-up by angular momentum loss) with $\vert\zeta\vert < 2\vert \Omega \vert$ and the \emph{Dedekind-like} (spinning-down) with $\vert \zeta \vert > 2 \vert \Omega \vert$ \cite{1993ApJS...88..205L, 1995ApJ...442..259L}. For the purposes pursued here, we address systems along the Jacobi-like sequence and, in virtue of its expected radiation signature, we call them \emph{chirping ellipsoids}, \CEL.

Paper \cite{1995ApJ...442..259L} studied the case of a newborn neutron star described by a polytropic index $n\leq1$. In that case, the spin-up sequence has a first chirp-like epoch (i.e. frequency and amplitude increase; see Fig.~7 in \cite{1995ApJ...442..259L}) and both the spin-down and spin-up epochs are in principle detectable by interferometers such as LIGO/Virgo \cite{1995ApJ...442..259L,2017ApJ...851L..16A}. 
No other values of the polytropic index, of interest e.g. for white dwarfs, have been explored in depth.

We follow the treatment of compressible ellipsoids by \cite{1993ApJS...88..205L,1994ApJ...420..811L} and refer the reader there for technical details. 
The dynamical timescale, and hence the unit of time, in our calculations will be $\tau_{\rm CEL} = 1/\sqrt{\pi G \bar{\rho}_0}$, where $\bar{\rho}_0$ is the mean density of the non-rotating star with the same polytropic index and total mass $M$, but with radius $R_0$ different from the mean radius $R=(a_1a_2a_3)^{1/3}$ of the compressible ellipsoid \cite{1993ApJS...88..205L}.

\joseprd{
When the polytropic index is close to $3$, the value of $\Omega/\sqrt{\pi G \bar{\rho}_0}$ along the equilibrium sequence is of the order of $10^{-2}$. With this information, we can infer the kind of astrophysical object whose GWs are within the frequency band of space-based detectors. For instance, for a GW frequency of the order of the minimum noise of \eLISA\, i.e. $f\sim 10$~mHz (see Fig.~\ref{fig:detectability} in Sec.~\ref{sec:4}), then  $\bar{\rho}_0\sim 5\times10^7$~g~cm$^{-3}$, which is a typical average density of a white dwarf (see, e.g., \cite{2011PhRvD..84h4007R}).
}

\section{GW emission of a \CEL}\label{sec:3}

In the weak-field, low-velocity approximation, the GW power of a rotating object is \cite{1975ctf..book.....L, maggiore2008gravitational, 2017grav.book.....M}
\begin{equation}
    \frac{dE}{dt} = -\frac{32}{5}\frac{G}{c^5}\Omega^{6}(I_{11} - I_{22})^2,
    \label{eqn:luminosity_gw}
\end{equation}
where $I_{ii} = \kappa_n M a_i^2/5$, with $\kappa_n$ a structure constant that depends on the polytropic index. \joseprd{We recall that the GW is quadrupole dominant and the angular frequency is twice the rotational one, i.e.~$\omega = 2\Omega$.} 

\textcolor{black}{
The GW amplitude
\begin{equation}
    h_0 = 4 \frac{G}{c^4 D} (\pi f)^2 (I_{11} - I_{22}),
\end{equation}
where $f$ is the GW frequency, and $D$ is the distance to the source and the typical GW emission timescale,
\begin{equation}
    \tau_{\rm GW} = \frac{f}{\dot{f}} = \Omega\frac{dE}{d\Omega}\frac{dt}{dE}, 
\end{equation}
are obtained from the equilibrium sequence of Riemann type-S ellipsoids described in Sec.~\ref{sec:2}; see Fig.~\ref{fig:hf_tau}.}

\textcolor{black}{
It can be seen from Fig.~\ref{fig:hf_tau} that these CELs can be considered as \emph{quasi-monochromatic}, i.e., $\tau_{\rm GW}\gg T_{\rm obs}$, where $T_{\rm obs}$ is the observing time of the space-based detector. This feature is very important to assess the detectability and degeneracy properties. Figure~\ref{fig:hf_tau} also shows that these spin-up CELs have a chirp-like early epoch, i.e. both the frequency and amplitude increase with time.
}

Different circulations converge during this early phase, characterized by axes ratios $\lambda_2=a_2/a_1, \lambda_3=a_3/a_1 \lesssim 0.7$. The smaller the polytropic index, the more deformed the star is during this chirping epoch. 

\begin{figure}
    \centering
    \includegraphics[width=0.7\textwidth]{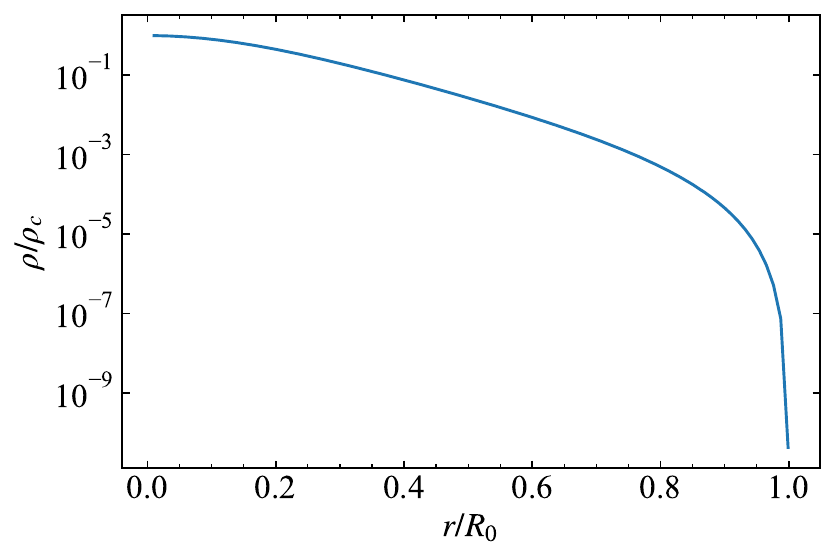}
    \caption{Density profile of a self-gravitating, non-rotating object whose internal matter is described by a polytropic equation of state with $n=2.95$}
    \label{fig:density_profile-nonrot}
\end{figure}
\begin{figure}
    \centering
    \includegraphics[width=0.49\textwidth]{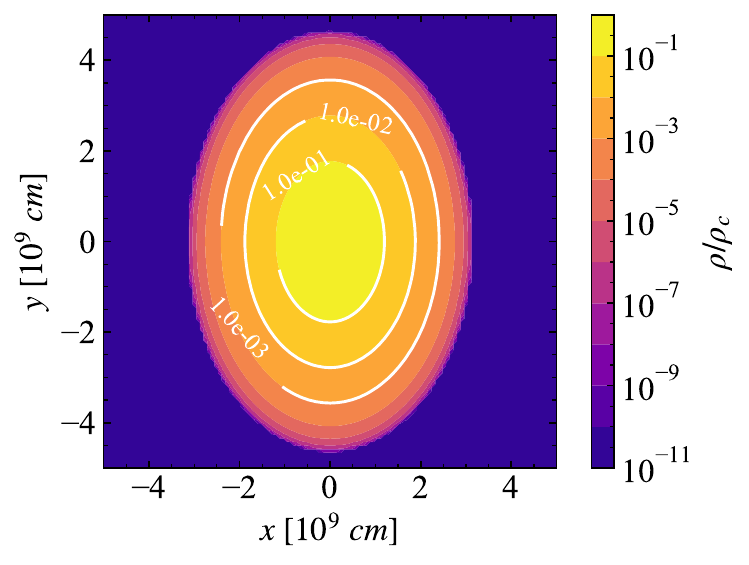}
    \includegraphics[width=0.49\textwidth]{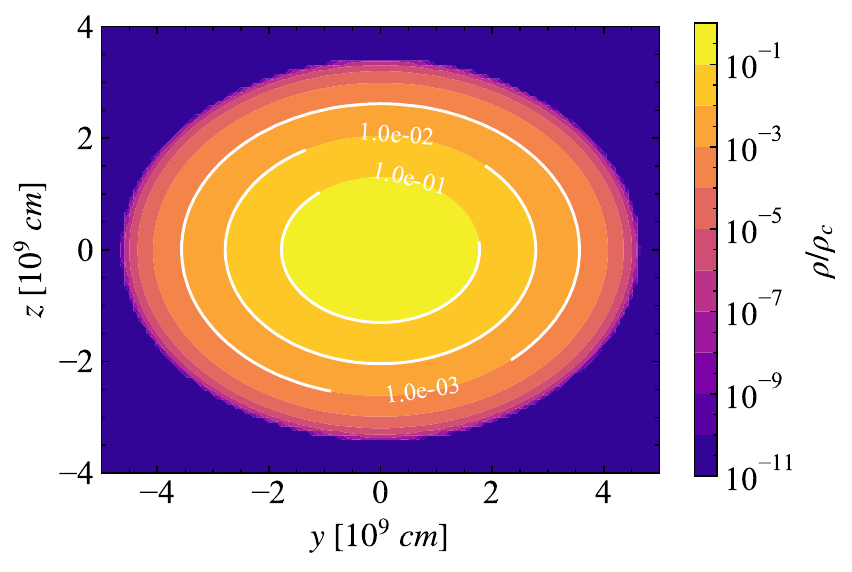}
    \caption{\textcolor{black}{Isodensity curves of a CEL with polytropic index $n=2.95$, $M_{\rm CEL}=1.2\,M_\odot$, and central density $\rho_c = 1.2\times10^8$~g~cm$^{-3}$, rotating with angular velocity $\Omega/\sqrt{\pi G \bar{\rho_0}}=0.037$. All the curves are \emph{self-similar} to the ellipsoid with axes ratio $a2/a_1 = 0.74$ and $a_3/a_1 = 0.80$. The compactness of the corresponding non-rotating configuration, with the same mass, is $\mathcal{C} = 0.44\times10^{-4}$.}}
    \label{fig:density-profile-CEL}
\end{figure}

We have identified CELs with deformed white dwarf-like objects. We show in Fig. \ref{fig:density-profile-CEL} isodensity contours of a CEL with polytropic index $n=2.95$, when the axes ratios are $\lambda_2 = 0.74, \lambda=0.80$. At this stage, the object rotates with angular velocity $\Omega/\sqrt{\pi G \bar{\rho_0}}=0.037$. All the contours represent self-similar ellipsoids, and the density profile is the same as the one of the non-rotating polytrope with the same index $n$ and radius $R_0$ \cite{1993ApJS...88..205L}, shown in Fig. \ref{fig:density_profile-nonrot}. At this point of the evolution $R/R_0 = 14.81$, which implies that for a CEL with $M=1.2 \Msun$ the axes are $a_1 = 7.1\times 10^9,a_2= 5.2\times 10^9, a_3= 5.6\times 10^9$ cm. When we compare with the simulations of double white dwarf mergers, evidently, there is a problem in scale that comes from the fact that the CEL is very expanded, nearly one order of magnitude compared with the non-rotating configuration. It is important to mention that this point of the CEL evolution is near the end of the chirp $\Omega_{\rm end
}/\sqrt{\pi G \bar{\rho_0}}\approx0.04$.

\joseprd{
We advance the possibility that these CELs might be the aftermath of double white dwarf mergers. Numerical simulations
(see, e.g., \cite{1990ApJ...348..647B, 2009A&A...500.1193L, 2012ApJ...746...62R, 2013ApJ...767..164Z, 2014MNRAS.438...14D}) have shown that the merged object is composed of a central white dwarf made of the undisrupted primary white dwarf and a corona made with nearly half of the disrupted secondary. The central remnant is surrounded by a Keplerian disk with a mass given by the rest of the disrupted secondary because very little mass ($\sim 10^{-3}~M_\odot$) is ejected during the merger.
}

\begin{figure}
    \centering
    \includegraphics[width=0.8\textwidth]{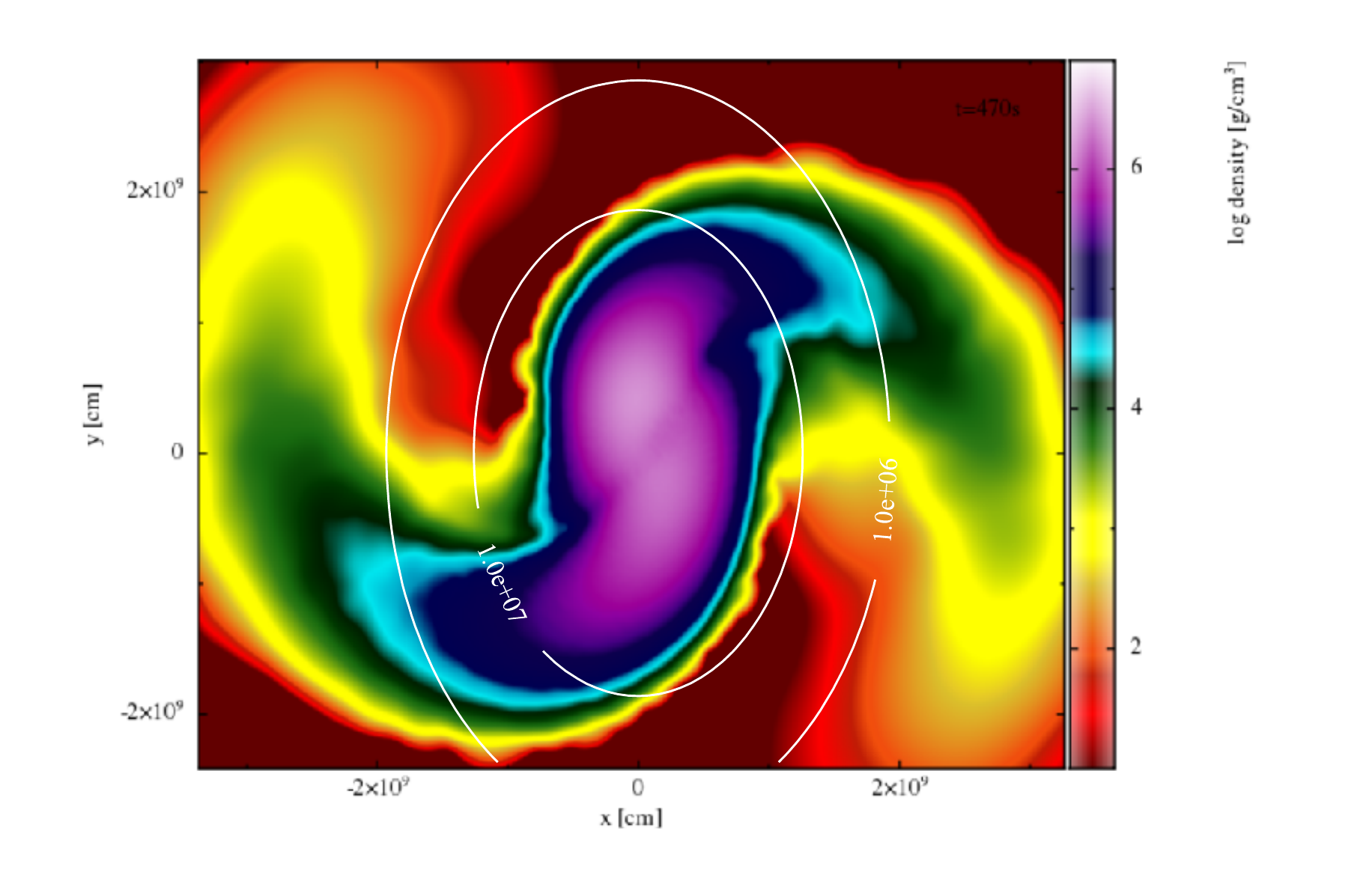}\\
    \includegraphics[width=0.8\textwidth]{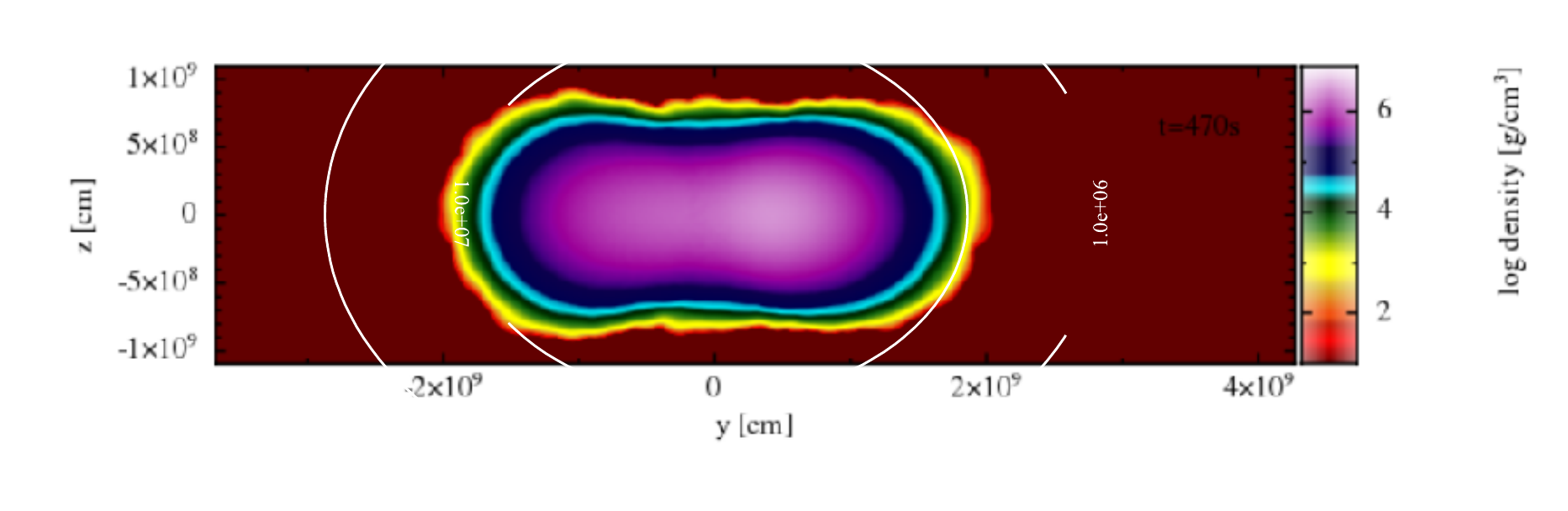}
    \caption{\textcolor{black}{Density map of a section in the orbital plane (top panel) and in the polar plane (bottom panel) of a $0.6+0.6~M_{\odot}$ double white dwarf merger simulated with $5\times 10^4$ SPH particles. This snapshot is taken $9$ orbital periods after mass transfer begins. The simulation was performed with an adapted version of PHANTOM \cite{2018PASA...35...31P}. This figure has been created using SPLASH (Price 2007), an SPH visualization tool publicly available at \url{http://users.monash.edu.au/~dprice/splash} \cite{2007PASA...24..159P}.}}
    \label{fig:simulation}
\end{figure}

\begin{figure}
    \centering
    \includegraphics[width=0.6\textwidth]{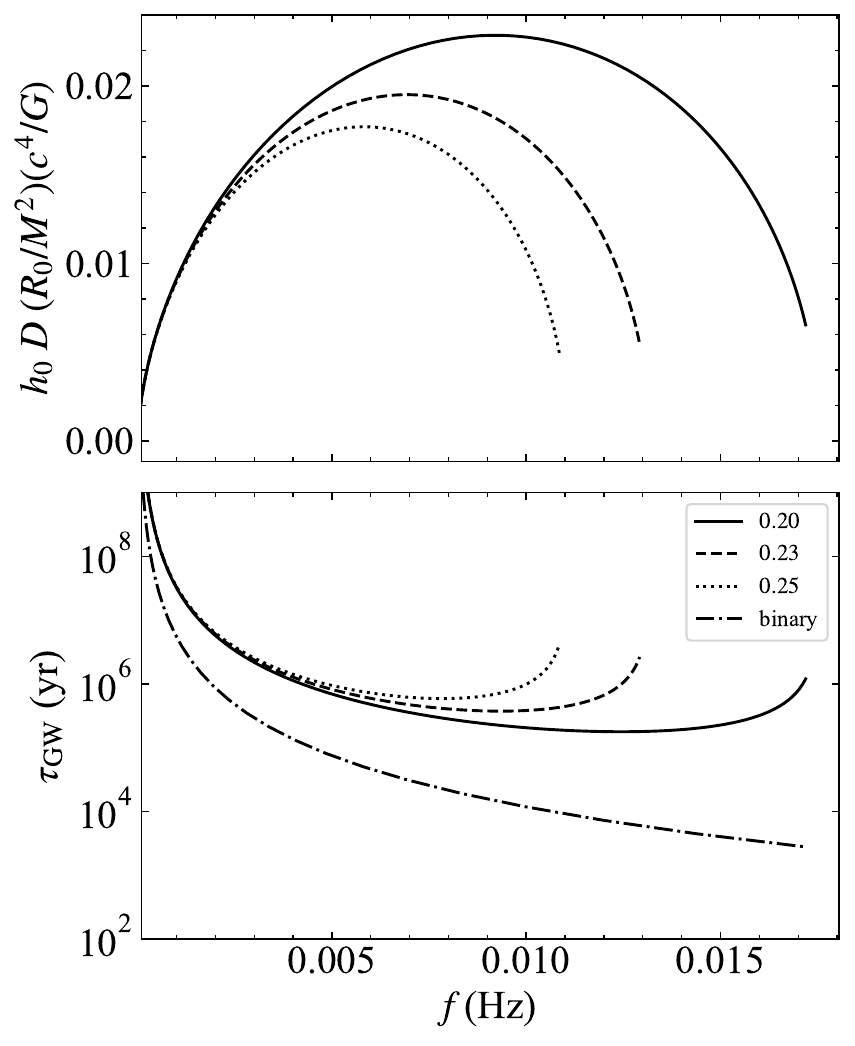}
    \caption{\textcolor{black}{GW amplitude (upper panel) and timescale (lower panel) as a function of the GW frequency (in Hz) for a compressible \CEL\ with polytropic index $n=2.95$. The \CEL\ mass is $M = 1.0\ M_\odot$, the same as the non-rotating spherical star with radius $R\approx 6000$~km. For comparison purposes, the dot-dashed lines show a binary with chirp mass equal to the \CEL\ mass. The systems are quasi-monochromatic, i.e.~$\tau_{\rm GW}\gg T_{\rm obs}$, at the frequency band of spaced-borne detectors.}}
    \label{fig:hf_tau}
\end{figure}

\joseprd{
To validate the above hypothesis, we have performed smoothed-particle-hydrodynamics (SPH) simulations of double white dwarf mergers to compare the structure of the post-merger, central white dwarf remnant with the one of the CEL. In Fig.~\ref{fig:simulation}, we show the density color map on the orbital and polar plane of a $0.6+0.6~M_{\odot}$ merger at about $9$ orbital periods after the starting time of the mass transfer. The two white dwarfs have merged, forming a newborn central white dwarf remnant. By comparing Figs.~ \ref{fig:density-profile-CEL} and \ref{fig:simulation}, we can conclude that the density and radii of the central white dwarf, the product of a merger, are similar to the ones of our relevant CELs, which validate our initial guess.
}

Turning to the comparison with a binary, we computed, as a first guess, the GW emission for a binary with total mass, $M_{\rm bin} = m_1+m_2$, and a chirp mass, $M_{\rm chirp} = M_{\rm bin} \nu^{3/5}$ ($\nu\approx q \equiv m_2/m_1$), equal to the mass of the \CEL. We found that the timescale and amplitude evolution of the binary is of the same order of magnitude as the ones of a \CEL. Hence, we conjecture that the two signals can have similar waveforms sweeping the same frequency interval simultaneously. 

\joseprd{
The GWs from these CELs, besides their early chirping-like behavior, are highly monochromatic. Hence, this poses a detection-degeneracy issue with other monochromatic systems, such as some kind of binaries, which we identify in Sec.~\ref{sec:5}.
}

For a more detailed and quantitative study of the waveforms, we used the non-dimensional parameter
\begin{equation}
    Q_{\omega} \equiv \frac{\omega^2}{\dot{\omega}}=  \frac{d\phi}{d\ln\omega}=  2\pi\frac{ d N}{d\ln f}=2\Omega^2\frac{dE}{d\Omega}\frac{dt}{dE} ,
    \label{eqn:Qomega}
\end{equation}
where $\phi$ is GW phase and $N$ the number of cycles. This parameter is an intrinsic measure of the phase-time evolution \cite{2013PhRvD..87h4035D}, and is (gauge) invariant under time and phase shifts. 

We make an empirical fit of $Q_\omega$ for \CEL\ with different indexes $n$, and different values of the compactness parameter $\mathcal{C} = G M_{\rm CEL}/(c^2 R_0)$. The fitting function is:
\begin{equation}
   Q_{\omega}^{\rm CEL}\approx \frac{ \mathcal{A}_n}{ \mathcal{C}^{5/2}}\biggl[\frac{\omega}{\sqrt{\pi G \bar{\rho}_0}}\biggr]^{\alpha}
    \label{eqn:Qw_fit}
\end{equation}
where the values of $\mathcal{A}_n$ and $\alpha$, as obtained fitting the waveforms of CELs with different polytropic structure constants, are shown in Table~\ref{tb:polyn}.

\begin{table}
\begin{center}
\begin{tabular}{ccccccc}
$n$ & $\kappa_n$ & $k_1$ & $k_2$ & $k_3$ & $\mathcal{A}_n$ & $\alpha$ \\
\hline
2.0 & 0.38712 & 1.1078 & 0.71618 & 1.6562 & 4.003 & $-1.222$ \\
2.5 & 0.27951 & 1.4295 & 0.67623 & 1.4202 & 4.060 & $-1.447$\\
2.7 & 0.24109 & 1.55971 & 0.66110 & 1.33194 & 5.926 & $-1.365$\\
2.9 & 0.20530 & 1.69038 & 0.64630 & 1.24621 & 4.940 & $-1.571$\\
2.95 & 0.19676 & 1.72309 & 0.64265 &1.22511 & 4.369 & $-1.614$ \\
2.97 & 0.19340 & 1.73617 & 0.64119 & 1.21669 & 3.760 & $-1.640$ \\
2.99 & 0.19005 & 1.74925 & 0.63973 & 1.20829 & 3.817 & $-1.652$
\end{tabular}
\caption{Polytropic structure constants ($n, \kappa_n, k_1, k_2, k_3$) and the $Q_{\omega}$ power-law empirical fitting parameters.}\label{tb:polyn}
\end{center}
\end{table}

The function $Q_\omega$ for both the \CEL\ and the binary has a power-law behavior but with a different exponent. The negative exponent implies that both have a monotonically increasing frequency. In the case of the \CEL, this behavior can be understood from the conservation of circulation and the interplay of compressibility and vorticity. Riemann S-type ellipsoids have internal motions with uniform vorticity contributing to the total angular momentum. In spin-up configurations, the radiation of angular momentum induces a vorticity loss. However, since the circulation is conserved, this loss must be compensated with a change in the angular velocity and the axes $a_1, a_2$. Thus, the spin-up of a \CEL\ has two ``components'': one due to the change in geometry that depends on the compressibility, and the other one due to the decrease of vorticity. The compressibility of the object changes with the polytropic index, inducing the behavior seen in Table \ref{tb:polyn} (e.g.~when $n\to3$ $\alpha\to -5/3$)\footnote{A similar behavior but for an \emph{axially symmetric} rotating star (Maclaurin spheroid) has been pointed out in \cite{1990ApJ...357L..17S}. There, it has been shown that when $n\to3$, the star can spin up by losing angular momentum (see also \cite{2018ApJ...857..134B} for a detailed analysis).}.

\begin{figure}
\centering
        \includegraphics[width=0.7\textwidth]{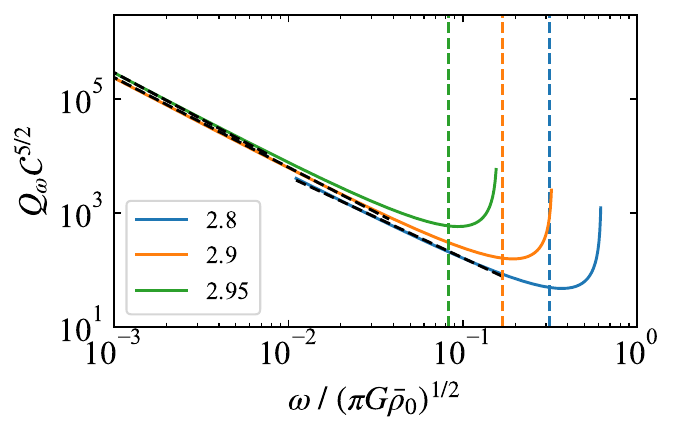}
\caption{Intrinsic phase-time evolution for a Riemann S-type spinning-up ellipsoid (\CEL) with polytropic indexes $n=2.8$, $2.9$, $2.95$ (blue, orange, green), normalized by the compactness parameter $\mathcal{C}$. Fits are shown as black dashed curves. The vertical dashed line represents the end of the chirping regime for each index, i.e. the frequency where the GW amplitude reaches the maximum. 
\label{fig:deltaq}}
\end{figure}

Empirical power-laws, such as that in Eq. \ref{eqn:Qw_fit}, can be used to compute analytically the phase-time evolution of the GW. The frequency and phase are the functions of $\tau = t_{\infty} - t$, where $t_{\infty}$ is the time the frequency formally diverges. \textcolor{black}{For binaries with point-like components, the GW frequency diverges when the orbital separation approaches zero.} In a real \CEL, this time is never achieved since the object ``leaves'' the chirping regime at a time $t_{\rm end}$ with a finite angular frequency $\omega_{\rm end}$ (see dashed lines in Fig.~\ref{fig:deltaq}).

\section{GW detectability}\label{sec:4}

\joseprd{
For our analysis, we assume that the matched filtering technique is used to analyze the GW data. In this case, the expected signal-to-noise ratio is given by (see e.g. \cite{1992PhRvD..46.5236F})
\begin{equation}
    \biggl(\frac{S}{N}\biggr)^2 = \langle \rho^2 \rangle = 2\times 4\int_{f_0}^{f_1} \frac{\langle |F_+\tilde{h}_+   + F_{\times} \tilde{h}_{\times}|^2 \rangle}{S_n(f)}df 
    \label{eqn:SNR}
\end{equation}
where $f_0$ and $f_1$ are the initial and final observed GW frequencies, $\tilde{h}_+(f),\tilde{h}_{\times}(f)$ are the Fourier transforms of the GW polarizations, $F_+, F_{\times}$ are the detector antenna patterns, and $S_n(f)$ is the power spectrum density of the detector noise. The factor $2$ comes from considering two Michelson interferometers ($6$ total laser beams).}

As a first approximation, the modulation of the projection onto the detector is estimated by performing an average over the source position and polarization angle. The inclination of the angular velocity with respect to the line of sight has also been averaged. The Fourier transform of the GW polarizations, $\tilde{h}_+$ and $\tilde{h}_{\times}$, can be obtained with the stationary phase method \cite{maggiore2008gravitational}. As usual, the characteristic amplitude is:
\begin{equation}
    h_c \equiv h_0 \sqrt{\frac{dN}{d\ln f}}    \;\stackrel{\text{opt}}{=}\; f \sqrt{2\bigl(|\tilde{h}_+|^2 + |\tilde{h}_{\times}|^2 \bigr)}, 
    \label{eqn:hchar}
\end{equation}
where the second identity is true \emph{only} when the \CEL\ is optimally oriented. The expected (angle averaged) signal-to-noise ratio is related to the latter characteristic amplitude by
\begin{equation}
    \langle \rho^2 \rangle = \frac{6}{25}\int_{f_0}^{f_1} \frac{h_c^2}{f^2 S_n(f)}df. \label{eqn:SNR-hc}
\end{equation}

\joseprd{
Since these \CEL\ are quasi-monochromatic, the expected signal-to-noise ratio can be readily estimated with the ``reduced'' characteristic amplitude,  $\tilde{h}_c$, defined as \cite{1998PhRvD..57.4535F}:
\begin{equation*}
    \tilde{h}_c (f) = h_0(f)\sqrt{N} = h_0(f)\sqrt{fT_{\rm obs}},
\end{equation*}
}
\textcolor{black}{
that applied to Eq.~(\ref{eqn:SNR-hc}) implies
\begin{equation}
    \langle \rho^2 \rangle \propto \frac{\tilde{h}_c^2(f_0)}{f_0 S_n(f_0)}.
\end{equation}
}

\begin{figure}
\centering
        \includegraphics[width=0.7\textwidth]{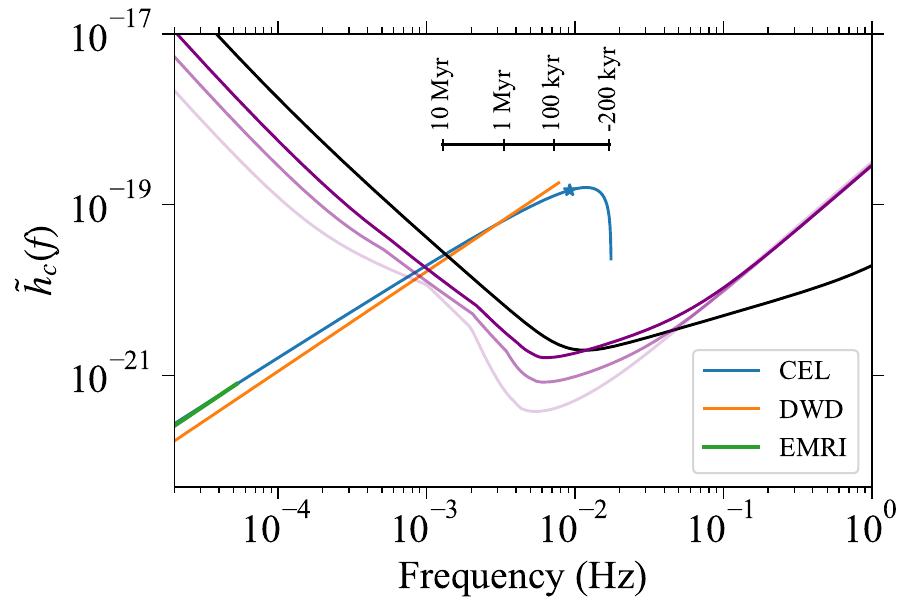}
\caption{Reduced characteristic amplitude,  $\tilde{h}_c$, of a \CEL, a double white dwarf, and an EMRI. The \CEL\ has a mass $M_{\rm CEL} =  1.0\ M_\odot$ and compactness $\mathcal{C}\approx 2.5\times10^{-4}$ (blue), according to the relativistic Feynman-Metropolis-Teller equation of state \cite{2011PhRvD..84h4007R}. The polytropic index is $n = 2.95$ and is located at a distance $D=1$~kpc. The observing time has been set to $T_{\rm obs} = 2$~yr. The blue dot at $f_{\rm end}\approx 9.20$~mHz marks the end of the chirping regime of the \CEL. The inset shows a chart with the time to reach the end of the chirp, $\tau_{\rm end}$. The EMRI is composed of $m_1 = 1940.62\ M_\odot$, $m_2 = 10^{-4}\ M_\odot$; it is located at $D=1.29$~kpc, and its evolution is shown up to the tidal-disruption frequency (green). The double white dwarf is composed of $m_1 = 0.45\;M_\odot$, $m_2 = 0.18\;M_\odot$; it is located at $D=1.20$~kpc, and its evolution is shown up to the point of Roche-lobe overflow (orange). For more details, see Table~\ref{tb:equiv}. Fits of the noise amplitude spectral density of \eLISA\ are shown as purple continuous lines with decreasing intensity for the configurations N2A1L4, N2A2L4, and N2A5L4, from top to bottom, respectively (see \cite{2016PhRvD..93b4003K} for the explicit form of the fits and conventions meaning). The amplitude spectral density of the \TianQin\ project detector is shown as a black continuous curve \cite{2016CQGra..33c5010L}.}
\label{fig:detectability}
\end{figure}

\textcolor{black}{
Figure~\ref{fig:detectability} shows $\tilde{h}_c $ for a \CEL\ with $n=2.95$ and $M_{\rm CEL} = 1.0\,M_\odot$. Furthermore, in order to illustrate the frequency vs. time evolution of the \CEL, we show in the same figure a panel with the time to reach the end of the chirping regime, $\tau_{\rm end} = t_{\rm end} - t$.  At $
\tau_{\rm end}=0$ this CEL reaches the GW frequency of $\approx 9.20$~mHz, after which the GW amplitude decreases.
}


\begin{landscape}
\begin{table}
\centering
\begin{tabular}{ccc|ccccc|cccc|c}
    $M_{\rm CEL}$ & $\mathcal{C}$  & $f_{\rm end}^{\rm CEL}$ & $\mchirp$  & $m_1$ &  $m_2$  &  $f_{\rm end}^{\rm bin}$& \multirow{2}{*}{Type-like} & $f_0$ &\multirow{2}{*}{$\Delta \phi_{\rm 1y}$} & \multirow{2}{*}{$\frac{\Delta h_0}{h_0}\big\vert_{\rm 1y}$}  & \multirow{2}{*}{ $\frac{D_{\rm CEL}}{D_{\rm bin}}$} & \multirow{2}{*}{SNR} \\ 
    $(M_{\odot})$ & $(10^{-4})$  & (mHz )& $(\Msun)$ & $(\Msun)$ & $(\Msun)$ & (mHz) &  &  (mHz) & & & &\\
    \hline
    1.0 & 2.5  & 9.20 & 0.32 & 1940.62 & 0.0001 & 0.053 & EMRI & 0.05 & $3.631\times 10^{-10}$ & $5.937\times10^{-13}$ & 0.778 & und. \\ 
        &    &        & 0.28 & 0.35 & 0.30 & 13.38 & PG1101+364 & 1.0 &$ 5.004\times10^{-5}$ & $2.515\times 10^{-9}$ & 0.773 & 0.687\\
        &    &        & 0.24& 0.45 & 0.18 & 7.76 & J0106-1003 & 3.0 &$5.018\times10^{-3}$ & $6.455\times 10^{-8}$ & 0.835 & 9.079 \\
    \hline
    1.4 & 20.0 & 148.70 & 0.48 & 2916.81 & 0.0015 & $0.064$ & EMRI & $0.05$ & $5.521\times10^{-10}$ & $9.322\times10^{-13}$& 0.808 & und. \\ 
        &      &       & 0.45 & 0.59 &  0.45 & 19.92 & WD0028-474 & 1.0 & $3.868\times 10^{-5}$ & $3.106\times10^{-9}$ & 0.776 & 1.511\\ 
        &      &       & 0.43& 0.52 &  0.47 & 21.30 & WD0135-052 & 3.0 & $2.660\times 10^{-3}$ & $6.344\times10^{-8}$ & 0.766 & 23.88\\ 
        &      &       & 0.42 & 0.51 &  0.45 & 20.25  & WD1204-450 & 6.0 & $4.148\times 10^{-2}$ & $4.377\times10^{-7}$ & 0.763 & 119.89\\ 
        &      &       & 0.41 & 0.47 & 0.47 & 21.48 & WD1704-481\footnote{Same chirp mass} & 9.0 & $2.135\times10^{-1} $ & $1.375\times10^{-6}$ & 0.764 & 145.73\\
\end{tabular}
\caption{\textcolor{black}{Parameters of a \CEL\ with $n=2.95$ and some of its equivalent binaries. The CEL is characterized by the mass and the compactness $\mathcal{C}$ obtained from the relativistic Feynman-Metropolis-Teller equation of state \cite{2011PhRvD..84h4007R}. We calculated the parameters of CEL so that there is a degeneracy with the corresponding binary. We emphasize that the degeneration is potentially subject to the naturalness of the existence of the triaxial object; see discussion below Eq. \eqref{eq:Rwdwd}. The frequency at the end of the chirping phase of the \CEL\ is denoted by $f_{\rm end}^{\rm CEL}$, i.e. when the CEL reaches its maximum GW amplitude. For each CEL, the chirp mass and the mass of the components of the equivalent binary are shown in the fourth, fifth, and sixth columns, respectively. The frequency at the end of the binary chirping regime, $f_{\rm end}^{\rm bin}$, is reported in the seventh column. This value is set, for the case of double white dwarfs, by the point when one of the stars reaches Roche-lobe overflow and, for the case of an EMRI, by the point of tidal disruption, assuming $R_2\approx 70,000$~km for the radius of the less massive component of the EMRI, $m_2$. The binary type is shown in the eighth column. When the system is a double white dwarf, the name of the most similar observed system is shown. For more details on the double white dwarfs, we refer the reader to \cite{2017MNRAS.466.1575R}, and references therein. The initial observing frequency $f_0$ is shown in the ninth column. The two systems' phase and relative amplitude differences after $1$~yr are shown in the next two columns. The ratio of the distance of both systems, assuming optimal orientations, is shown in the twelfth column. The corresponding signal-to-noise ratio of the CEL, located at $D_{\rm CEL}=1$~kpc, for an observing time of $1$~yr from when $f=f_0$ is shown in the last column.}}\label{tb:equiv}
\end{table}
\end{landscape}

\textcolor{black}{
For a distance between the detector and the source of 1~kpc used in Fig.~\ref{fig:detectability}}, $\tilde{h}_c$ is well above the \eLISA\ noise curve, at least near the end of the chirping regime, so the GW is in principle detectable. The typical value of $\omega/(\pi G \bar{\rho}_0)^{1/2}$ during the chirping phase is $\sim 10^{-5}$--$10^{-1}$. For typical densities of a white dwarf $\sim 10^6$--$10^{9}$~g~cm$^{-3}$, the frequency is $\sim 10^{-6}$--$10$~Hz, inside the \eLISA\ sensitivity band. The detectability properties obtained from Eqs.~\eqref{eqn:hchar} and \eqref{eqn:SNR-hc} are reported in the last column of Table~\ref{tb:equiv}.

\textcolor{black}{
In addition, the \CEL\ can be regarded as monochromatic in some part of their lifetime. Figure~\ref{fig:deltaq} shows that the evolution is rather slow at low frequencies, and becomes slower when $n \to 3$. Thus, the \CEL\ is expected to be monochromatic in those regions.
} 

\begin{figure}
    \centering
    \includegraphics[width=0.65\textwidth]{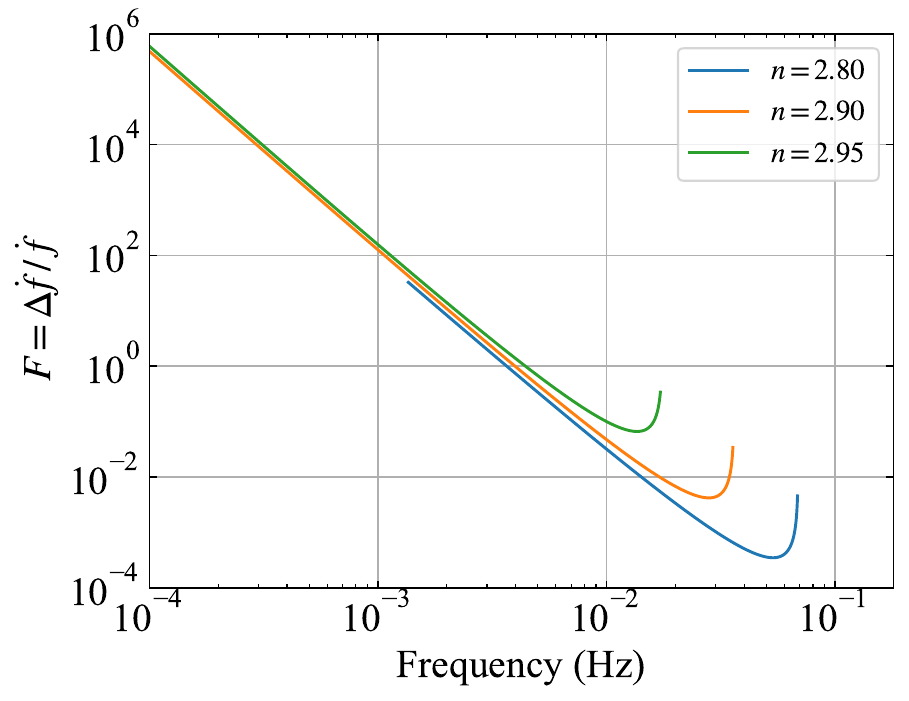}
    \caption{\joseprd{Ratio of $\Delta \dot{f}$, error in estimating the time derivative of the frequency, to the value $\dot{f}$ of CELs with different polytropic indices, $F=\Delta  \dot{f}/\dot{f}$. The ratio was obtained assuming $(S/N)=10$ and $T_{\rm obs}=2$~yr. When $F>1$, the error in estimating the frequency is larger than the theoretical value of the \CEL, i.e., the time derivative of $f$ is inside the error, and the system can be regarded as monochromatic \cite{2002ApJ...575.1030T}. For $f\lesssim 3$~mHz, CELs are monochromatic for the adopted detection value.}}
    \label{fig:monochromaticity}
\end{figure}

\joseprd{
Specifically, whether a GW is monochromatic depends on the detector's frequency resolution or frequency bin, $T_{\rm obs}^{-1}$, on the signal-to-noise ratio, and the frequency evolution of the \CEL. The errors in estimating the frequency and its change rate by matched filtering are \cite{2002ApJ...575.1030T}
\begin{align}
     \Delta f &= 0.22\biggl(\frac{\langle \rho \rangle}{10}\biggr)^{-1} T_{\rm obs }^{-1},\\
     \Delta \dot{f} &= 0.43\biggl(\frac{\langle \rho \rangle}{10}\biggr)^{-1} T_{\rm obs}^{-2},\label{eqn:Delta_fdot}
\end{align}
which are frequency independent for $T_{\rm obs}\gtrsim 2$~yr. The ratio of the error in $\dot{f}$, to the rate of change of the frequency of a \CEL\ can be used to determine its ``monochromaticity'' \cite{2002ApJ...575.1030T}, i.e.
\begin{equation}
     F \equiv \frac{\Delta \dot{f}}{\dot{f}}.
    \label{eqn:monochronomacy}
\end{equation}
Thus, a source can be assumed as monochromatic for the detector if $F>1$. We show this criterion for different polytropic indices in Fig.~\ref{fig:monochromaticity}, from which it is confirmed that in some parts of the sensitivity band, the CELs are monochromatic.
}

\textcolor{black}{
Summarizing, our estimates indicate that CELs are detectable for $1$~yr of observation (see Table~\ref{tb:equiv}), given they have appreciable deformation and are close enough, $D\lesssim$ 1~kpc. Detectability depends also on the frequency. The system is monochromatic at very low frequencies, $f<1$~mHz. Still, its GW amplitude (at $D=1$~kpc) is not high enough to accumulate sufficient signal-to-noise ratio in $1$~yr to be detected (see Table~\ref{tb:equiv} and Fig.~\ref{fig:detectability}).
}

\section{CEL-binary degeneracy} \label{sec:5}

We now compare the above results with the ones associated with specific binary systems. In the quasi-circular orbit approximation, the intrinsic phase-time parameter of a binary has a power-law exponent equal to $-5/3$. For a \CEL\ whose equation of state is modeled as an ultra-relativistic degenerate electron gas ($n=3$), the intrinsic phase has the same exponent as the binary, \joseprd{which confirms our initial hypothesis}. Therefore, there exists a binary system, with an appropriate value of the chirp mass, that matches the phase-time evolution of the CEL \joseprd{(see below).} 

When $\alpha = -5/3$, the dependence on the compactness in Eq.~\eqref{eqn:Qw_fit} disappears. 
\textcolor{black}{It is interesting that this behavior finds a simple physical explanation in a compact star such as a white dwarf: the ultra-relativistic limit for a Newtonian self-gravitating star made of fermions is approached when $\rho \to \infty$, namely when $R\to 0$. In this limit, the star properties become radius-independent when the critical mass is reached.
}

\begin{figure}
    \centering
    \includegraphics[width=0.65\textwidth]{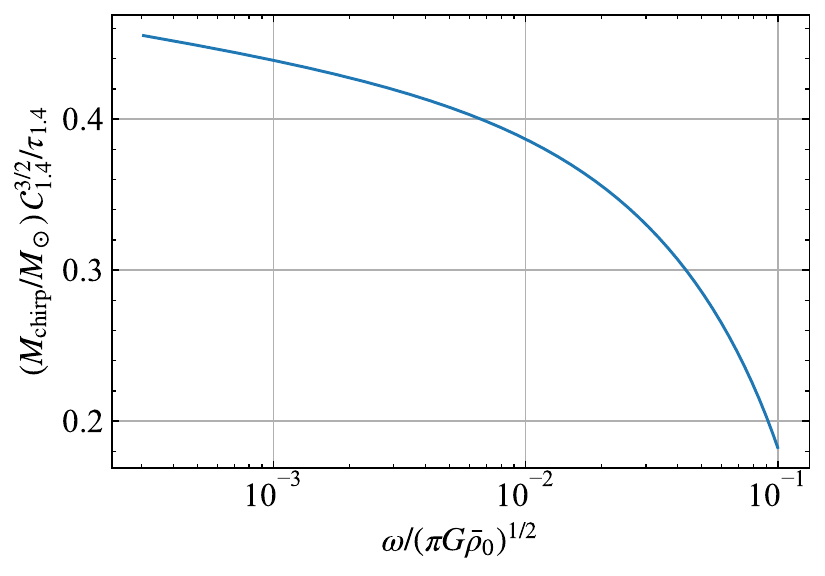}
    \caption{\joseprd{Chirp mass of the binary with the same intrinsic phase-time evolution $Q_{\omega}$ of a CEL of $n=2.95$, at a GW angular frequency $\omega$. The value has been normalized by $\mathcal{C}_{1.4}\equiv \mathcal{C}/(2\times10^{-3})$ and $\tau_{1.4}\equiv (\pi G \bar{\rho}_0/\bar{\rho}_{1.4})^{-1/2}$, where $\bar{\rho}_{1.4}$ is the mean density of a non-rotating white dwarf with mass $1.4\,M_\odot$ and radius $R_{\rm WD} \approx 1000$~km, according to the mass-radius relation obtained from the relativistic Feynman-Metropolis-Teller equation of state \cite{2011PhRvD..84h4007R}. Therefore, the values shown in this plot correspond to a CEL with $M_{\rm CEL}=1.4\,M_\odot$, $\mathcal{C}=2\times10^{-3}$.}}
    \label{fig:mchirp_equiv}
\end{figure}

\joseprd{
For each \CEL\ at a given frequency, a binary system with the same intrinsic phase-time evolution parameter $Q_{\omega}$ exists. Hereafter, we illustrate the analysis with a CEL whose polytropic index is close to 3, ie. $n=2.95$. 
It can be seen that at $\omega/\sqrt{\pi G \bar{\rho}_0}\sim10^{-3}$ the chirp mass is $\sim 0.4\ M_{\odot}$ and scales with the compactness, $\mathcal{C}_{1.4}^{3/2}$, where $\mathcal{C}_{1.4} \equiv \mathcal{C}/(2\times10^{-3})$ (see Figure~\ref{fig:mchirp_equiv}).
} 

\begin{figure}
    \centering
    \includegraphics[width=0.65\textwidth]{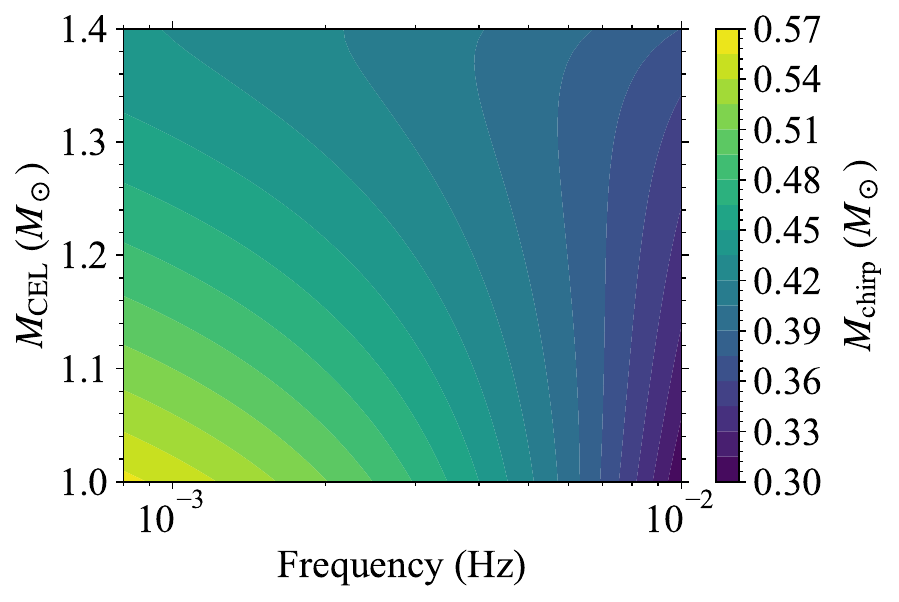}
    \caption{\textcolor{black}{Contours of constant chirp mass of the equivalent binary as a function of the \CEL\ mass and the observed frequency. In general, the chirp mass of the equivalent binary depends on the ${\cal C}$, $M_{\rm CEL}$, and the observed frequency $f$. However, once the equation of state is selected, the mass-radius relation is fixed, implying that $M_{\rm chirp}$ depends only on $M_{\rm CEL}$ and $f$. We have here used for the white dwarf the ``Chandrasekhar'' equation of state (see text for details).}}
    \label{fig:Mcel2Mchirp}
\end{figure}

\textcolor{black}{
To give a complete vision of the CEL-binary degeneracy, we show in Fig.~\ref{fig:Mcel2Mchirp} the chirp mass of the equivalent binary as a function of the observed frequency and the mass of the CEL ($n=2.95$). The mass-radius relation of the non-rotating white dwarf-like object has been obtained for a Chandrasekhar-like equation of state, i.e. the pressure is given by the electron degeneracy pressure while the density is given by the nuclei rest-mass density.\footnote{\textcolor{black}{Differences in the white dwarf mass-radius relation for more general equation of state are negligible for the scope of this work (see, e.g., \cite{2011PhRvD..84h4007R})}}.
}

\begin{figure}
    \centering
    \includegraphics[width=0.65\textwidth]{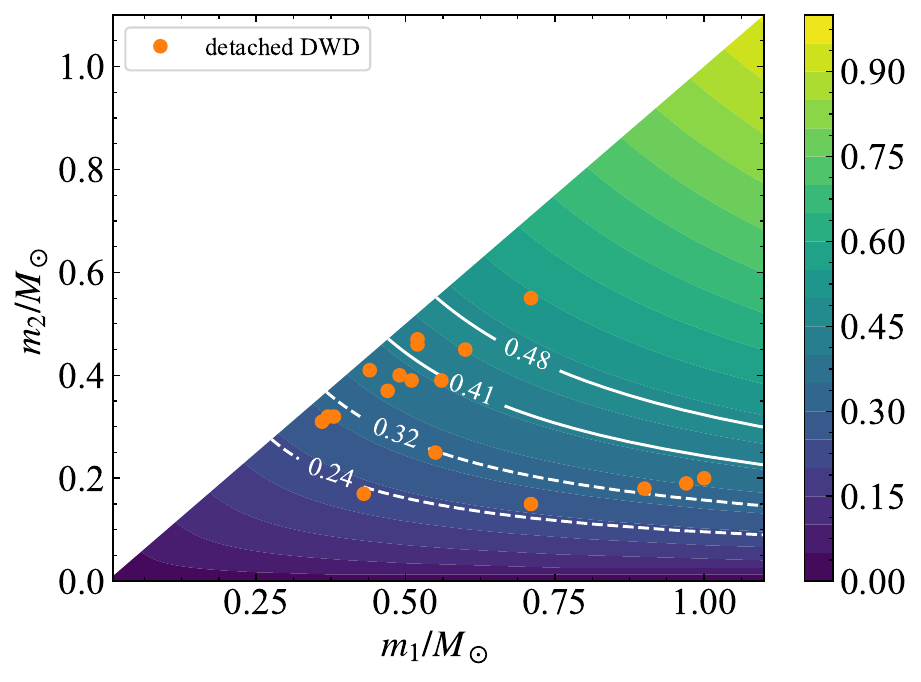}
    \caption{\joseprd{Contours of constant chirp mass of binaries in the mass range of double white dwarfs. Observed detached double white dwarfs with measured parameters \cite{2017MNRAS.466.1575R} are shown as orange circles. The continuous white contour lines correspond a binary with $M_{\rm chirp}=(0.41, 0.48)\,M_\odot$, which matches the $Q_{\omega}$ of a CEL with $\mathcal{C}=2\times 10^{-3}$, at $f = (9.00, 0.05)$~mHz, respectively. The dashed white contour lines correspond to a binary with $M_{\rm chirp}=(0.24, 0.32)\,M_\odot$, which matches the $Q_{\omega}$ of a CEL with $\mathcal{C}=2.5\times 10^{-4}$, at $f=(3.00, 0.05)$~mHz, respectively.}}
    \label{fig:m1m2_chirp}
\end{figure}

\textcolor{black}{
For a given chirp mass, there is a degeneracy in the masses of the binary components, i.e. there exist many combinations of $m_1$ and $m_2$ produce CEL equivalent binaries (see Fig.~\ref{fig:m1m2_chirp}). Here, we focus on two types of equivalent binaries: 1) detached double white dwarfs and 2) EMRIs composed of an intermediate-mass black hole and a planet-like object. It is worth mentioning that the chirp mass of observed detached double white dwarfs, with the currently measured parameters, ranges from $0.23$ to $0.61\,M_\odot$ \cite{2017MNRAS.466.1575R}. For illustration purposes, we calculated some equivalent binaries to a CEL ($n=2.95$), and show the results in Table~\ref{tb:equiv}.
}

\joseprd{\emph{In the limit $n\to 3$}, the following relation must be satisfied to have identical phase-time evolution:
\begin{equation}
    \frac{3}{5} 2^{7/3} \mathcal{A}_{3}\biggl(\frac{3}{4}\biggr)^{5/6} = \biggl(\frac{M_{\rm CEL}}{M_{\rm bin}}\biggr)^{5/3}\frac{1}{\nu}.
    \label{eqn:cond}
\end{equation}}
\joseprd{The right-hand side of the last expression is of the order of $(M_{\rm CEL}/M_{\rm bin})^{5/3}\nu^{-1}\approx 10$ (see Table \ref{tb:polyn}). Consequently, when the chirp mass, $M_{\rm chrip} = M_{\rm bin} \nu^{3/5}$, and the mass of the \CEL\ are of the same order, both waveforms have the same phase-time evolution. Equation \eqref{eqn:cond}, can be used to estimate readily the equivalent chirp mass. It is worth mentioning that in the actual calculation, we used the intrinsic phase-time parameter given by the numerical solution of the Riemann S-type sequence and not the one given by the fit.}

\begin{figure}
    \centering
    \includegraphics[width=0.75\textwidth]{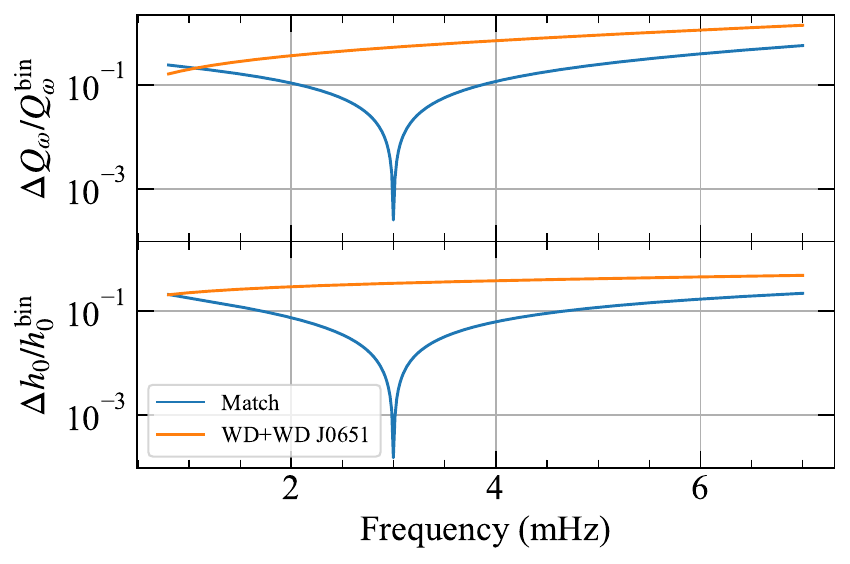}
    \caption{\textcolor{black}{Residuals of comparison of the GW phase (top panel) and amplitude (bottom panel), of a CEL ($n=2.95$) with $M_{\rm CEL}=1.0\,M_\odot$ and $\mathcal{C}=2.5\times10^{-4}$, and an equivalent binary system with $M_{\rm bin}=0.24\,M_\odot$. The binary matches the $Q_{\omega}$ of the CEL at $f=3$~mHz (dip in the residuals). For reference, we also show in orange, the residual of comparing the CEL signal with that of the detached double white dwarf J0651 ($M_{\rm chirp}=0.31\,M_\odot$). The compared binaries are located at the same distance from the detector, while the ratio of the distances to the EMRI and the CEL is $D_{\rm bin}/D_{\rm CEL} = 1.2$.}}
    \label{fig:Q-DeltaQ}
\end{figure}

\joseprd{ When the chirp mass has been matched, the two systems have nearly equal phase-time evolution and are, in practice, indistinguishable in their phases. This feature can be appreciated in Fig.~\ref{fig:Q-DeltaQ}, where we compare and contrast the intrinsic phase-time evolution of a CEL and binary systems with matching and non-matching (but close) chirp mass. Some LISA targets that do not match the phase-time evolution of a CEL are an EMRI composed of a massive black hole, e.g. $m_1 = 10^5\,M_\odot$ and $m_2 =1\,M_\odot$, or a binary neutron star, e.g. $m_1=m_2=1.3 \,M_\odot$. However, a double white dwarf (also a known LISA target) like J0651, currently the second shortest orbital period known GW emitter \textcolor{black}{in the mHz frequency band \cite{2012ApJ...757L..21H}}, has a chirp mass close to the matching one; thus, its phase-time evolution around $1$~mHz is nearly equal to the CEL under consideration (see Fig.~\ref{fig:Q-DeltaQ}).
}

\joseprd{
It could be argued that the signal match is not exact for the range of frequencies considered. However, it must be noticed that, since both systems are quasi-monochromatic, differences between the evolution parameters appear when the frequency changes appreciably. They become out of phase only when the observation is performed over very long periods of time $\gg 4$~yr.
}

\joseprd{
We now estimate how much the systems get out of phase by integrating $\Delta Q_{\omega} = |Q_{\omega}^{\rm CEL} - Q_{\omega}^{\rm bin}|$, during $1$~yr, i.e.
\begin{equation}
    \Delta \phi_{\rm 1y} = \int_{\omega_0}^{\omega_{\rm 1yr}}\Delta Q_{\omega} d\ln \omega, 
    \label{eqn:Delta_phi}
\end{equation}
where $\omega_0$ is the initial observed GW angular frequency and $\omega_{\rm 1yr}$ is the GW angular frequency after $1$~yr. The results are presented in Table~\ref{tb:equiv}. As observed, phase differences are extremely small for most of the considered values of $M_{\rm CEL}$ and $n$. The systems (CEL and binary) are monochromatic at very low frequencies and show full degeneracy.}

\joseprd{
Regarding the GW amplitude, we found that $h_c \propto f^{-1/5}$ and this holds almost for any $n$.  Therefore, the reference amplitude $h_0$ scales as} $h_0 \propto f^{-1/5 - \alpha/2}$.

\joseprd{Although in the limit $n\to 3$ the intrinsic phase-time evolution of the \CEL\ and the binary tend to follow the same power-law exponent, the \CEL\ amplitude $h_0 \propto f^{0.63}$ grows with a different (but nearly equal) exponent. For example, once some chirp mass has matched the phase, the distance to the source can be chosen to match the GW amplitudes. We have found that the distances must be of the same order. Again, since the exponents are nearly equal and the evolution during observing time is slow, the GW amplitudes remain nearly equal, as shown in the examples of Table~\ref{tb:equiv} and Fig.~\ref{fig:Q-DeltaQ}.
}

\textcolor{black}{
The end of the chirp regime for a binary depends on its nature. For the case of a double white dwarf, this is generally given by the Roche-lobe overflow. Thus, we set this frequency using the Eggleton approximate formula for the effective Robe-lobe radius \cite{1983ApJ...268..368E}. The radius of each component has been obtained assuming a polytropic equation of state with $n=1.5$ \cite{1967aits.book.....C}, since in this case, the matching binary has low-mass components. Roche-lobe overflow frequencies for selected double white dwarfs are reported in Table~\ref{tb:equiv}. For the case of an EMRI, the limit is due to the tidal disruption of the less massive component. The GW frequency at tidal disruption is:
$$f_{\rm td}\approx (G m_2/R_2^3)^{1/2}/(2.4^{3/2}\pi),
$$
where $R_2$ is the radius of the (less massive) component $m_2$, and the tidal radius is $r_{\rm td}\approx 2.4q^{-1/3}R_2$ \cite{chandrasekharellipsoidal} (see also Table~\ref{tb:equiv}).
}

\joseprd{
The above detection degeneracy might be broken since the chirping phase of the CEL and the binary, owing to Roche-lobe overflow or tidal disruption, end at different frequencies. It would then be possible to discriminate between systems by observing above some frequency. For instance, if the observation is carried out near and beyond the Roche-lobe overflow frequency, the continuation of a chirping power-law with exponent $\approx -5/3$, will point to a CEL ($n=2.95$). In contrast, if the power-law changes, this will hint at the possibility that the system is a double white dwarf that just filled one of its Roche-lobes. In addition, degeneracy between an EMRI and a CEL is broken, owing to the fact the former can not be \emph{individually} detected by currently planned space-based detectors (see Table~\ref{tb:equiv}).
}

\joseprd{
Finally, in the low-velocity, weak-field limit, any monochromatic GW can be \emph{considered} as being radiated from a deformed (not axially symmetric) rotating star. Equivalently, any monochromatic GW can be thought as produced from a circular binary. The correspondence between monochromatic GWs and sources is not one-to-one. The appropriate identification of the source (if possible) relies on the astrophysical implications of the characterizing parameters and/or additional astronomical data, such as the relative abundance of the two systems.
}

\bigskip

In summary, \textcolor{black}{the above results show that given a CEL \emph{with $n$ close to 3}, a binary system can be found whose \emph{GW chirping evolution during observing times} matches the one of the CEL, and vice-versa. When this chirping evolution is not identifiable due to the slow intrinsic evolution, short periods of observation, or both, the system's true nature would be highly uncertain.} 

\joseprd{As already stated, CELs can be monochromatic. Thus, detection degeneracy extends to even more systems. Namely, in the monochromatic regime, there is a degeneracy between CELs, or between CELs and binaries with parameters different from previously found. This kind of degeneracy will be addressed elsewhere.} 

\section{Rate of equivalent binaries and \CEL}\label{sec:6}

Next, to assess the impact of the binary-\CEL\ detection-degeneracy, we estimate the rate of both sources in the local universe for the sensitivity of \eLISA\ at the frequencies of interest (Fig.~\ref{fig:detectability}). 
We adopt the source parameters of Table \ref{tb:equiv}. 

The equivalent EMRIs found for the \CEL\ 
are formed by an intermediate-mass black hole with a mass in the range $m_1=500$--$3000\;\Msun$ and a substellar, planet-like object $m_2\approx\nu m_1=(0.7$--$4)\times 10^{-3}\Msun$. The latter mass range corresponds approximately to masses between the one of Saturn ($M_{\rm Sat}=3\times 10^{-4}\;\Msun$) and the one of Jupiter ($M_{\rm Jup}\sim 10^{-3}\;\Msun$). Intermediate-mass black holes in this mass range have been suggested by observations and simulations, at least for the case of dynamically, old globular clusters (see e.g.~\cite{2008ApJ...681.1431M}). It has also been suggested that dwarf spheroidal galaxies may harbor intermediate-mass black holes in their cores (see e.g.~\cite{2015PhRvD..92b2002G}). In the latter case, however, the galaxy core may also be explainable as a dark matter concentration alternative to the intermediate-mass black hole \cite{arguelles2018novel}.

Even if the association of intermediate-mass black holes with planetary-mass objects is absent in the literature, extensive work has been done testing different dynamical processes 
in the core of young stellar clusters, able (at least numerically) to drive the formation of intermediate mass-ratio binary inspirals (IMRIs).  These IMRIs typically include an intermediate-mass black hole and a compact stellar object (stellar-mass black holes, neutron stars, or main sequence stars). These results, at least for our purposes,
can shed some light on the odd systems here considered. 

We assume that those dynamical mechanisms operate independently of the mass of the captured compact object (obeying the equivalence principle). If this assumption is true, the challenge is whether planetary-sized objects could be found at the core of globular clusters and dwarf spheroidals.

Planetary formation in globular clusters has been a matter of debate for several decades \cite{1992ApJ...399L..95S,Davies2001,Beer2004,2007MNRAS.381..334S}.  
Still, and against all odds, to the date of writing, at least one planet has been discovered in the globular cluster M4. The planet has a similar Jupiter-like mass and orbits a binary system formed by the millisecond pulsar PSR B1620–-26 and a white dwarf \cite{2000ApJ...528..336F}. More intriguingly, the system is located close to the cluster core, where the dynamical lifetime of planetary systems is much lower than the estimated binary age. This suggests that, at least in this case, the planet originally formed around its host star (the white dwarf's progenitor) while being far from the center. The star and its planet (or its entire planetary system) then migrate towards the center, encountering in the process the pulsar. Once there, the system may become unstable in a timescale of $10^{8}$~yr (see Eq.~5 in \cite{2000ApJ...528..336F}), and the planet will probably be detached from the system and eventually captured by the intermediate-mass black hole. \textcolor{black}{The details of this process will also depend on the complex dynamics of the cluster \cite{Meylan1997,Baumgardt2003,Fregeau2003}.} 

Let us assume that a fraction $f$ of the stars in the outer regions of a dynamically evolved globular cluster have planetary companions, and a fraction $g$ of them migrates towards the center in a multi-Gyr timescale. Further, we assume that once there, most systems become unstable, and the intermediate-mass black hole captures planets. Under these conditions, the formation rate of EMRIs at a given globular cluster can be estimated as $10^{-9} f\cdot g\cdot N_\star$~yr$^{-1}$. The number of globular clusters in the local Universe is uncertain, but it can be estimated within the local group (which occupies a volume of $4$~Mpc$^{-3}$). The Milky Way contain around $200$ (see e.g. \cite{2018ApJ...860L..27C} and references there in); Andromeda has the largest number with $460\pm70$ \cite{2001AJ....122.2458B}; M33 has only $30$ \cite{1991ARA&A..29..543H}; while the Large Magellanic Cloud has around $13$ \cite{2004AJ....127..897V}. Assuming at least $1$ globular cluster in the $\sim30$ dwarf galaxies of the local group, the total number of globular clusters within $1$~Mpc will be $\sim 10^3$. If only a fraction $\alpha$ of them contain an intermediate-mass black hole with a mass as large as that able to mimic the signal of a \CEL, namely $10^3\;\Msun$, the rate of EMRIs will be $R_{\rm EMRI}=10^{-6}\alpha\cdot f\cdot g\cdot N_\star$~yr$^{-1}$. 
Assuming 
$N_\star\sim 10^6$, $\alpha=0.2$--$1$, $f=0.5$, $g=0.1$--$1$, this rate 
becomes:
\begin{equation}\label{eq:Remri2}
R_{\rm EMRI}=0.02-0.5\;{\rm yr}^{-1}.
\end{equation}

\textcolor{black}{Another family of the identified equivalent binary systems corresponds to double white dwarfs. Since we are interested in the systems that can enter the interferometer frequency band, we now adopt double white dwarfs that can merge within the Hubble time. The merger rate of these systems in a typical galaxy is estimated to be $(1$--$80)\times 10^{-13}$~yr$^{-1}$~$M_\odot^{-1}$ (at $2\sigma$) 
\cite{2017MNRAS.467.1414M,2018MNRAS.476.2584M}. Thus, using $M=6.4\times 10^{10}~M_\odot$ for the Milky Way  \cite{2001ApJ...556..340K}, we obtain :
\begin{equation}\label{eq:DWDrate}
    R_{\rm DWD} = 0.0064-0.512\;{\rm yr}^{-1}.
\end{equation} 
}

Turning to the \CEL, we have seen that their structure (mass, radii, compactness, equation of state, etc.) points to a white dwarf-like nature. Deformed white dwarfs can result from mass transfer from a companion, see e.g. \cite{2010MNRAS.406.2749L}. The rate at which these events occur might be close to that of novae, which has been estimated in the Milky Way to be $\sim 10$--$80$~yr$^{-1}$ \cite{chen2016modelling} and, more recently, $\sim 27$--$81$~yr$^{-1}$ \cite{2017ApJ...834..196S}. 
If we assume that a fraction $\beta$ of all white dwarfs potentially becoming novae undergone a spin-up transition, the \CEL\ rate may be as high as:
\begin{equation}
R_{\rm CEL}=(10-80)\beta \;{\rm yr}^{-1}.
\end{equation}
Another, possibly more plausible mechanism for the formation of highly-deformed white dwarfs is the merging of double white dwarfs. Numerical simulations show that, when the merger does not lead to a type Ia supernova, the merged configuration is made of three regions \cite{1990ApJ...348..647B, 2004A&A...413..257G, 2009A&A...500.1193L,2012A&A...542A.117L,2012ApJ...746...62R,
2013ApJ...767..164Z, 2014MNRAS.438...14D}: a rigidly rotating, central white dwarf, on top of which there is a hot, differentially-rotating, convective corona, surrounded by a Keplerian disk. The corona comprises about half of the mass of the totally disrupted secondary star, while the rest of the secondary mass belongs to the disk since a small mass ($\sim 10^{-3}\,M_\odot$) is ejected. The rigid core$+$corona configuration has a structure that resembles our \CEL\ or triaxial object after the chirping regime (see Fig.~\ref{fig:simulation}). Depending on the merging components masses, the central remnant can be a massive ($1.0$--$1.5~M_\odot$), fast rotating ($P=1$--$10$~s) white dwarf \cite{2013ApJ...772L..24R,2018ApJ...857..134B}.

\textcolor{black}{
We adopt the view that the deformed white dwarfs result from double white dwarf mergers that do not lead to type Ia supernovae since the latter should lead to total disruption of the merged remnant, see Fig. \ref{fig:simulation}. We estimate the merger rate as the rate (\ref{eq:DWDrate}), subtracted off the type Ia supernova rate that is about $(12$--$22)\%$ of it \citep{2009ApJ...699.2026R}. Therefore, by requiring the double white dwarf merger channel to cover the supernova Ia population, we obtain a lower limit to the rate of deformed white dwarfs from such mergers, potentially observable as a \CEL\ within the Milky Way. Thus, we obtain:
\begin{equation}\label{eq:Rwdwd}
R_{\rm CEL}=\gamma(\epsilon)(0.0056-0.45)\;{\rm yr}^{-1}
\end{equation}
}
Since only eccentric mergers give rise to the final ellipsoidal-shape object, see e.g., \cite{2015MNRAS.450.2948A}, $\gamma(\epsilon)$ is a parameter indicating that fraction. Given that not all Riemann-S ellipsoids behave as a CEL, but only those with appreciable deformation (the chirping nature occurs at the beginning of the evolution, see Fig. \ref{fig:deltaq}), $\gamma$ is a function the ellipticity, $\epsilon \equiv (I_{11}-I_{22})/(I_{11}+I_{22})$. As far as we know, this parameter has not been obtained from simulations or observations. The possible observation of GW radiation from CELs or EM (see, e.g. \cite{2022ApJ...941...28S}) could constrain this parameter. On the other hand, this rate can be very similar to the EMRI rate estimated before (see Eq.~\ref{eq:Remri2}). Using an extrapolating factor of Milky Way equivalent galaxies, whose volume is $0.016$~Mpc$^{-3}$ \cite{2001ApJ...556..340K}, the above rate implies a local cosmic rate of $(0.74$--$5.94)\times 10^6$~Gpc$^{-3}$~yr$^{-1}$.
 

Therefore, we found that EMRIs, double white dwarfs, and CELs (here identified as deformed white dwarfs) could be numerous. The rates of CELs, as a function of the ellipticity, could be comparable to the ones of EMRIs and double white dwarfs.

\textcolor{black}{
Although the above rate of EMRIs is as high as that of the double white dwarf mergers or that of the \CEL, they do not represent an important source of degeneracy since the signal-to-noise ratio in one-year time of observation is very low, impeding their detection as \emph{single sources} by GW detectors (see Table~\ref{tb:equiv})}. \joseprd{However, given their very likely high occurrence rate, they might represent an important source of GW stochastic background, which will be studied elsewhere.}

Under these conditions, the CEL-double white dwarf potential degeneracy is a significant problem. The unambiguous identification of these sources would need to pinpoint its sky position and/or be able to observe above the frequency of Roche-lobe overflow of the less massive white dwarf in the double white dwarf system (see Fig.~\ref{fig:detectability} and Table \ref{tb:equiv}). Whether or not this would be achievable by the planned space-based facilities GW-detection remains a question to be answered. Still, it can be done via joint electromagnetic observations or future arrays of space-based interferometers.


\section{Conclusions}\label{sec:7}

\textcolor{black}{
Compressible, Riemann S-type ellipsoids with a polytropic index $n\gtrsim 2.7$, that we have called CELs, emit quasi-monochromatic GWs with a frequency that falls in the sensitivity band of planned space-based detectors (eg.~\eLISA\ and \TianQin; see Fig.~\ref{fig:hf_tau}). Inside the sensitivity band, CELs evolve sufficiently slowly to remain quasi-monochromatic during the planned observation times. These sources exhibit a chirp-like behavior similar to binary systems. In the limit $n \to 3$, as inferred from empirical fits shown in Table~\ref{tb:polyn}, both systems have the same intrinsic phase-time evolution $Q_{\omega}$. This behavior is due to the change in the compressibility of the CEL with $n$.
}

CELs located at galactic distances are detectable by planned space-based detectors during one year of observation (see last column of Table~\ref{tb:equiv}). We refer to CELs as those triaxial objects with \emph{appreciable} deformation, so they exhibit a chirping nature. We have found that within the detectors sensitivity band, a CEL ($ 2.9\lesssim n < 3.0$) having intrinsic, quasi-monochromatic parameters, $h_0, f, \dot{f}$, or equivalently $h_0, Q_{\omega}$ can have the same values of those of a binary, see Fig.~\ref{fig:mchirp_equiv} and Table~\ref{tb:equiv}. Namely, given a quasi-monochromatic binary characterized by its frequency, chirp mass, and distance, it can be found a CEL mass and distance, whose waveform at the same frequency has the same $\dot{f}$ (or $Q_{\omega}$) and amplitude of the binary. In this sense, CEL and quasi-monochromatic binaries could be degenerated, given the naturalness of CELs' existence to be determined. We have here pinpointed two kinds of quasi-monochromatic binaries potentially degenerated with CELs: double white dwarfs and EMRIs composed of an intermediate-mass black hole and a planet-like object.

\textcolor{black}{
The completely different physical nature of CELs and such binaries should allow, in principle, to distinguish them. Following this reasoning, we have found that the final frequency of the quasi-monochromatic chirping behavior of a binary is set, in the case of EMRIs (intermediate-mass black hole-planet), by the tidal disruption, or in the case of a double white dwarf, by Roche-lobe overflow. The tidal disruption frequency of this system is $\sim 10^{-5}$~Hz. These EMRIs cannot be detected as single sources by space-based detectors since they do not accumulate enough signal-to-noise ratio in the observing time (see Table~\ref{tb:equiv}). Thus, CELs and EMRIs do not pose the problem of detection degeneracy. For the systems considered in this work, the following relation is in general satisfied: $f_{\rm td}<f_{\rm RLOF}<f_{\rm end}^{\rm CEL}$. Consequently, observing a quasi-monochromatic GW (with chirp mass $\sim 0.5\,M_\odot$) above the Roche-lobe overflow frequency would strongly indicate a CEL. Below frequencies $\sim 10^{-2}$~Hz, the CELs and binaries are degenerated and cannot be distinguished using only GW data. In those cases, the electromagnetic data will be crucial in determining the real nature of the GW source.
}

\textcolor{black}{
Because of the relevance of this result for space-based detectors, we have discussed the current estimates of the occurrence rate of this kind of system. For the deformed white dwarfs, we adopted the view that they can be formed either by accretion from a companion or by double white dwarf mergers (see Fig.~\ref{fig:simulation}). Surprisingly, we found that rates of EMRIs, double white dwarf, and CELs could be comparable (see the discussion below Eq. \eqref{eq:Rwdwd}). Although EMRIs cannot be individually resolved, their occurrence rate makes them a plausible stochastic GW source that deserves a detailed analysis. However, this issue is beyond the scope of the present article and will be addressed elsewhere. From the present first approach, we can conclude that there might be a potential  GW source \textit{confusion}, for individually resolved events in the frequency range $f \lesssim 10$~mHz, between double white dwarfs and CELs. Despite this issue, it is possible to do science with these sources. Indeed, we have presented some possible solutions for the detection-degeneracy problem and encourage the scientific community to explore additional ones.}

\bigskip
\section*{Acknowledgments}
JMBL thanks support from the FPU fellowship by Ministerio de Educaci\'on Cultura y Deporte from Spain. J.F.R. thanks financial support from the Patrimonio Aut\'onomo - Fondo Nacional
de Financiamiento para la Ciencia, la Tecnolog\'ia y la
Innovaci\'on Francisco Jos\'e de Caldas (MINCIENCIAS -
COLOMBIA) under the grant No. 110685269447 RC-
80740–465–2020, project 69553, and from the Universidad Industrial de Santander, VIE, Contrato de financiamiento RC No. 003-1598/Registro Contractual 2023000357.

\bibliographystyle{apsrev}
\bibliography{references}

\end{document}